\newcommand\IPA{IPA-CuCl$_3$}
\documentclass[twocolumn,prl,floatfix,preprintnumbers,amsmath,amssymb,superscriptaddress]{revtex4}

\usepackage{graphicx}
\usepackage{dcolumn}
\usepackage{bm}

\begin{document}

\title{Excitations from a Bose-Einstein condensate of magnons in coupled spin ladders.}

\author{V. O. Garlea}
\affiliation{Neutron Scattering Sciences Division, Oak Ridge
National Laboratory, Oak Ridge, Tennessee 37831-6393, USA.}

\author{A. Zheludev}
\affiliation{Neutron Scattering Sciences Division, Oak Ridge
National Laboratory, Oak Ridge, Tennessee 37831-6393, USA.}

\author{T. Masuda}
\affiliation{International Graduate School of Arts and Sciences,
Yokohama City University, 22-2, Seto, Kanazawa-ku, Yokohama City,
Kanagawa, 236-0027, Japan.}

\author{H. Manaka}
\affiliation{Graduate School of Science and Engineering, Kagoshima
University, Korimoto, Kagoshima 890-0065, Japan.}

\author{L.-P.~Regnault}
\author{E.~Ressouche}
\author{B.~Grenier}
\affiliation{CEA-Grenoble, DRFMC-SPSMS-MDN, 17 rue des Martyrs,
38054 Grenoble Cedex 9, France.}

\author{J.-H. Chung}
\altaffiliation{ Department of Materials Science and Engineering,
University of Maryland, College Park, Maryland, 20742, USA}
\affiliation{NCNR, National Institute of Standards and Technology,
Gaithersburg, Maryland 20899, USA.}

\author{Y. Qiu}
\altaffiliation{ Department of Materials Science and Engineering,
University of Maryland, College Park, Maryland, 20742, USA}
\affiliation{NCNR, National Institute of Standards and Technology,
Gaithersburg, Maryland 20899, USA.}

\author{K. Habicht}
\author{K. Kiefer}
\affiliation{BENSC, Hahn-Meitner Institut, D-14109 Berlin,
Germany.}

\author{M. Boehm}
\affiliation{Institut Laue Langevin, 6 rue J. Horowitz, 38042
Grenoble Cedex 9, France.}

\date{\today}

\begin{abstract}
The weakly coupled quasi-one-dimensional spin ladder compound
(CH$_3$)$_2$CHNH$_3$CuCl$_3$ is studied by neutron scattering in
magnetic fields exceeding the critical field of Bose-Einstein
condensation of magnons. Commensurate long-range order and the
associated Goldstone mode are detected and found to be similar to
those in reference spin-dimer materials. However, for the upper
two massive magnon branches the observed behavior is totally
different, culminating in a drastic collapse of excitation
bandwidth beyond the transition point.
\end{abstract}

\maketitle

Bose-Einstein condensation (BEC), such as the superfluid
transition in liquid $^4$He \cite{London1938}, is the emergence of
a collective quantum ground state in a system of interacting
Bosons. The condensate is characterized by a macroscopic order
parameter that spontaneously breaks a continuous $U(1)$ symmetry.
For BEC to occur at $T>0$, the Bosons should be able to freely
propagate in 3 dimensions (3D) \cite{LLIX}. In one dimension BEC
is forbidden even at zero temperature. In a striking example of
dimensional crossover, even weak 3D coupling can enable BEC in 1D
systems, where the normal state itself results from the unique 1D
topology. A realization of this peculiar quasi-1D case was
proposed only recently \cite{Giamarchi1999}, and involves the
condensation of magnetic quasiparticles in weakly coupled
antiferromagnetic (AF) spin ladders.

Magnetic BEC can occur in a variety of spin systems
\cite{Batyev1984}. In gapped quantum magnets, for example, an
external magnetic field drives the energy of low-lying magnons to
zero by virtue of Zeeman effect, prompting them to condense at
some critical field $H_c$ \cite{Affleck91}. This transition is
fully equivalent to conventional BEC. The rotational $O(2)\equiv
U(1)$ symmetry is spontaneously broken by the emerging AF long
range order. At $T=0$ the density of magnons is zero for $H<H_c$,
and vanishingly small just above the transition. The phenomenon is
therefore described in the limit of negligible quasiparticle
interactions. To date, such transitions were mainly studied in
materials composed of coupled structural spin clusters
\cite{Shiramura1997,Kodama2002,Ruegg2003,Jaime2004,Sebastian2005,Zapf2006}.
The condensing quasiparticles are then local triplet excitation
that propagate due to inter-cluster interactions
\cite{Nohadani2004}. This is in contrast to the original model of
Ref.~\cite{Giamarchi1999} that deals with coupled extended,
translationally invariant objects. Their disordered ``spin
liquid'' normal state is a direct consequence of 1D topology
\cite{Haldane,Kennedy92}. Even for weak coupling the magnons are
fully mobile in 1D, rather than localized. Is the physics of the
field-induced BEC in quantum AF spin ladders any different from
that in local-cluster spin systems?

\begin{figure}
\includegraphics[width=8.7cm]{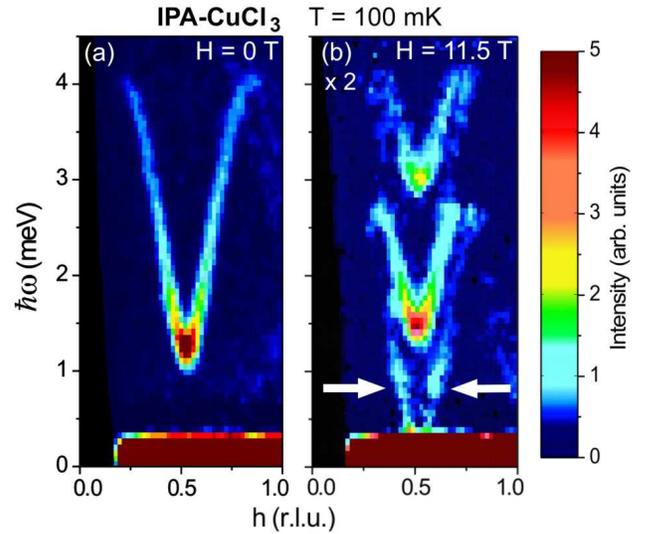}
\caption{Time-of-flight neutron spectra measured in \IPA\ in the
 1D Haldane-gap spin liquid phase (a) and the 3D ordered BEC phase (b), at $H=0$ and $H=11.5$~T,
 respectively.  Solid white
 arrows indicate the linearly dispersive gapless Goldstone mode.
 \label{Fig1}}
\end{figure}

In the present work we address this issue experimentally, through
a neutron scattering investigation of a prototypical spin ladder
material (CH$_3$)$_2$CHNH$_3$CuCl$_3$ (\IPA).  This compound
almost exactly realizes the original theoretical model of Ref.
\cite{Giamarchi1999}. Its  spin ladders are built of magnetic
$S=1/2$ Cu$^{2+}$ ions and run parallel to the $a$ axis of the
triclinic P$\overline{1}$ crystal structure. Conveniently, each
one can be viewed as a ``composite'' Haldane spin chain
\cite{Masuda2006}: pairs of $S=1/2$ spins on each rung are
strongly {\it ferromagnetically} (FM) correlated and act as
effective $S=1$ objects \cite{Masuda2006,Manaka2000}. Coupling
along the legs of the ladders is antiferromagnetic (AF), and
translates into AF interactions between effective spins in the
``composite'' $S=1$ chains. Such chains are gapped \cite{Haldane}
and are spin liquids with only short-range correlation. For \IPA\
the energy gap is $\Delta=1.2$~meV \cite{Manaka97,Masuda2006}.

\begin{figure}
 \includegraphics[width=8.5cm]{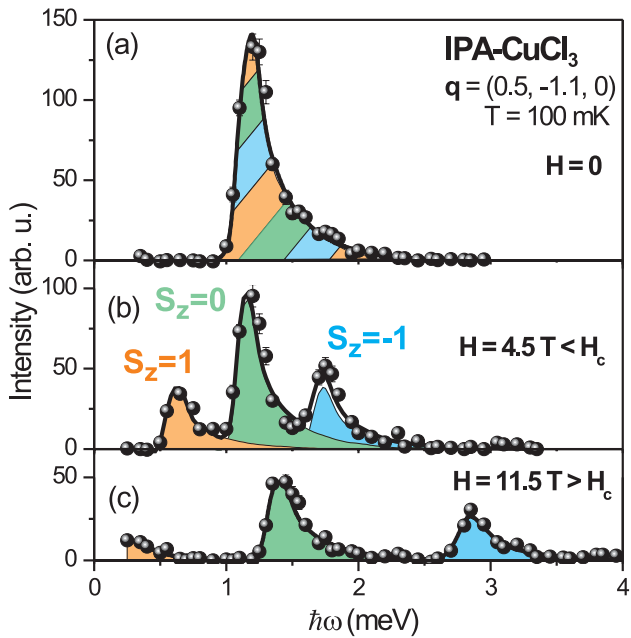}
 \caption{(Color online) Background-subtracted neutron spectra measured in \IPA\ at the 1D AF
  zone-center $h=0.5$ in various applied magnetic fields (symbols). The field values
  correspond to the 1D Haldane-gap spin liquid phase (a,b)
  and the 3D ordered magnon BEC phase (c). Line shapes are entirely
  due to experimental resolution (solid lines).
  \label{Fig2}
 }
\end{figure}

\begin{figure}
 \includegraphics[width=8.5cm]{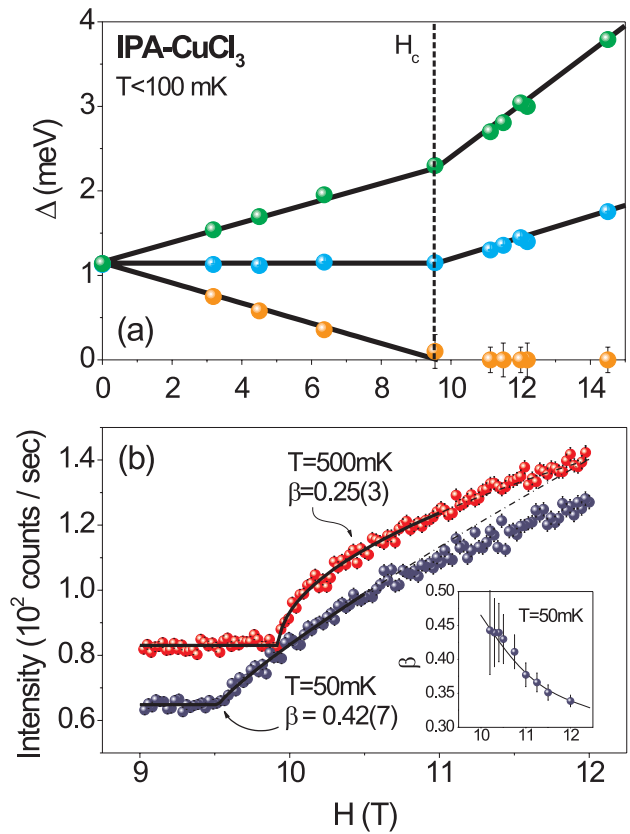}
 \caption{(Color online) (a) Measured gap energies in \IPA\ as a function of applied magnetic
  field.  (b) Measured field dependencies of the $(0.5,0,0)$ magnetic Bragg reflection at two
  different temperatures (symbols). The $y$-offset corresponds to the actual background for $T=50$~mK
  and is arbitrary for the $T=500$~mK data. Lines are power-law fits over
  a range of 1~T. Inset: critical exponent $\beta$ as a function
  of the field window used in the least squares fit at $T=50$~mK.
  \label{Fig3}
 }
\end{figure}

Excitations from this quantum-disordered ground state are revealed
in inelastic neutron scattering (INS) experiments that directly
probe the pair spin correlation function $S(\mathbf{q},\omega)$.
Fig.~\ref{Fig1}a shows a time-of-flight (TOF) spectrum collected
on a 3~g deuterated \IPA\ single crystal sample at $T=100$~mK in
zero magnetic field using the Disc Chopper Spectrometer (DCS) at
NCNR and 6.68~meV fixed incident energy neutrons. The magnon, with
a steep parabolic dispersion along the $a$ axis and gap at the 1D
AF zone-center $h= 0.5$, is clearly visible. $\Delta$ is small
compared to the chain-axis magnon bandwidth, but is considerably
larger than transverse bandwidths along the $c$ ($0.4$~meV) and
$b$ axes ($<0.1$~meV) \cite{Masuda2006}. Since our main purpose
will be to understand the special role that the AF spin ladder
structure plays in the BEC phase of \IPA, we shall be comparing
our results to those found in literature for TlCuCl$_3$
\cite{Ruegg2003}, a prototypical AF spin-dimer compound. There
$\Delta$ is small compared to magnon dispersion bandwidths in all
3 directions \cite{Cavadini2001,Oosawa2002}. To date, TlCuCl$_3$
is the only material for which the spectrum of spin excitations in
the BEC phase has been measured experimentally.

Figure~\ref{Fig2} illustrates the effect of magnetic field on
\IPA. It shows spectra collected  at the 1D AF zone-center $h=0.5$
in several fields using the SPINS 3-axis spectrometer at NCNR,
with 3.7~meV fixed-incident energy neutrons, focusing Pyrolitic
graphite analyzer and a BeO filter after the sample. As the field
is turned on, the single peak at $H=0$ (Fig.~\ref{Fig2}a) becomes
divided into three equidistant components (Fig.~\ref{Fig2}b). The
peak widths are resolution-limited. The measured field
dependencies of the gaps are plotted in Fig.~\ref{Fig3}a, which
also includes points obtained using the cold neutron 3-axis
spectrometer FLEX at HMI. The gap in the lower mode extrapolates
to zero at $H_c=9.6$~T, where a BEC of magnons was previously
detected in bulk measurements \cite{Manaka98}.

As the gap softens, commensurate long-range AF order sets in and
gives rise to new magnetic Bragg reflections of type
$(h+1/2,k,l)$, $h$, $k$ and $l$-integer. The magnetic structure
was determined at $H=12$~T in a neutron diffraction experiment at
the D23 lifting counter diffractometer at ILL, on a
3$\times$2$\times$9~mm$^3$ single crystal sample, using
$\lambda=1.276$~\AA\ neutrons. A good fit to 48 independent
magnetic reflections measured at $T=50$~mK was obtained using a
collinear model with spins perpendicular to the field, aligned
parallel to each other on the rungs, and antiparallel along the
legs of the ladders. The refined value of the ordered moment is
0.49(1)~$\mu_\mathrm{B}$.

The field dependence of the $(0.5,-1,0)$ peak intensity measured
at $T=50$~mK is plotted in Fig.~\ref{Fig3}b, lower curve. To
estimate the order parameter critical exponent $\beta$ we
performed power-law fits to the data in a progressively shrinking
field window (Fig.~\ref{Fig3}b, insert). The extrapolated value is
$\beta>0.45$, in agreement with expectations. Indeed, at
$T\rightarrow 0$, due to vanishing magnon density, one should
recover the mean field (MF) result $\beta=0.5$
\cite{Giamarchi1999,Matsumoto2002}. Any discrepancies between the
observed and MF behavior become more pronounced at elevated $T$,
when magnon density increases, and their interactions become
relevant. At $T=500$~mK, for example, in \IPA\ we get
$\beta=0.25(3)$ (Fig.~\ref{Fig3}b, upper curve). BEC critical
indexes have also been observed under appropriate conditions in
the dimer compound BaCuSi$_2$O$_6$ \cite{Sebastian2005}. However,
recent work showed that the BEC universality of the transition in
TlCuCl$_3$  is compromised by deviations from the Heisenberg
model. Anisotropy\cite{Sirker2005} and magnetoelastic
coupling\cite{Johannsen2005} modify the critical indexes and
account for a small gap in the ordered phase. To date, in \IPA\ we
found no evidence of lattice distortions at $H_c$ or deviations
from BEC behavior.

\begin{figure}
 \includegraphics[width=8.5cm]{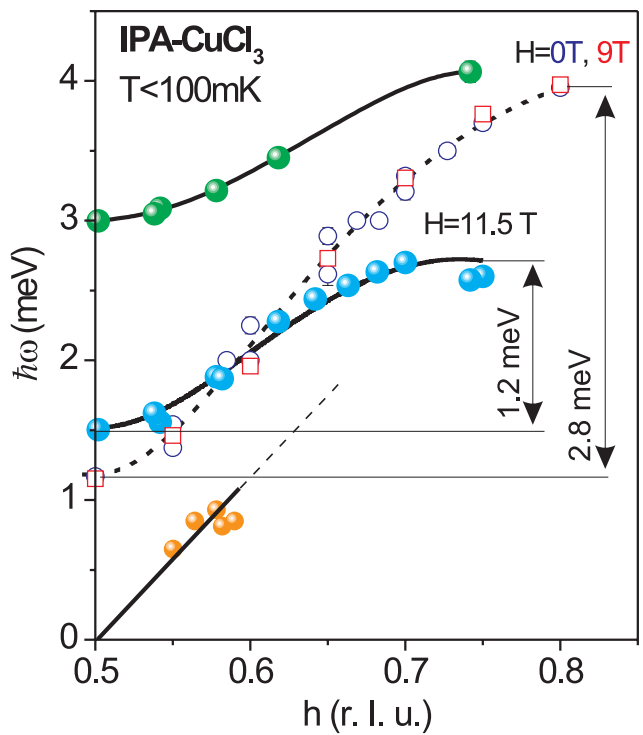}
 \caption{(Color online) Dispersion relation of the three excitation
 branches measured in \IPA\ at $H=11.5$~T$>H_\mathrm{c}=9.7$~T (solid
 symbols).
 Open circles: magnon dispersion at $H=0$ \protect\cite{Masuda2006}
 Open squares: Dispersion of the middle branch at $H=9$~T.
 Lines are guides for the eye.
  \label{Fig4}
 }
\end{figure}

A key result of this work is a direct measurement of excitations
of the magnetic Bose-Einstein condensate in the high-field phase.
Data collected above the critical field are shown in
Fig.~\ref{Fig2}c (SPINS) and in Fig.~\ref{Fig1}b (DCS). Three
distinct excitation branches, two gapped and one gapless, are
clearly visible, though the overall inelastic intensity is reduced
compared to lower fields. For each mode the dispersion relations
at $H=11.5$~T were obtained by fitting Gaussian profiles to
constant-$h$ cuts through the data in Fig.~\ref{Fig1}b. The
results are plotted in Fig.~\ref{Fig4} in solid spheres. For
comparison, we also plot the dispersion relation of the triplet at
$H=0$ (open circles) \cite{Masuda2006}. Open squares show the
dispersion of the middle magnon branch measured just below the
transition at $H=9$~T, on the IN14 3-axis spectrometer at ILL
under similar conditions.

The gapless mode observed at $H>H_c$ at low energies (arrows in
Fig.~\ref{Fig1}b) is fully analogous to the phonon in conventional
BEC, being the Goldstone mode associated with the spontaneous
breaking of $O(2)$ symmetry. It is a {\it collective} excitation
of the magnon condensate and has a linear dispersion relation
\cite{Matsumoto2002}. It takes the place of the massive
quadratically dispersive {\it single-magnon} excitation below
$H_c$ (Fig.~\ref{Fig1}a). At $H=11.5$~T the fitted velocity of the
Goldstone mode $v_\mathrm{G}=1.74(3)$~meV is reduced compared to
the spin wave velocity \footnote{For an excitation with gap
$\Delta$ we define the velocity as $v={d \sqrt{(\hbar
\omega)^2-\Delta^2}}/{d (2 \pi h)}$, to be measured in energy
units.} in zero field $v=2.9(2)$~meV \cite{Masuda2006}. This
behavior is qualitatively similar to that in TlCuCl$_3$
\cite{Ruegg2003,Matsumoto2002}, if one neglects the tiny
anisotropy gap in the latter system \cite{Sirker2005}, which is
too small to be detected with INS anyway. The similarity also
extends to the field dependence of the gap energies in those
magnon branches that do not soften at the transition point. In
both compounds the corresponding slope increases abruptly at $H_c$
(Fig.~\ref{Fig3}a). A bond-operator theoretical treatment of the
dimer model \cite{Matsumoto2002} attributes all these effects to
an admixture of the higher-energy triplet modes to the condensate.
This interpretation can be qualitatively extended to our case of
coupled spin ladders.

While the long-wavelength spectral features in \IPA\ are very
similar to those in simple spin-dimer systems, the {\it
short-wavelength} spin dynamics at the zone boundary is strikingly
different. We find that the two massive excitations in \IPA\
undergo a qualitative change upon the BEC transition. As seen in
Fig.~\ref{Fig1}b and Fig.~\ref{Fig4}, at $H=11.5$~T, only 20\%
above $H_\mathrm{c}$, their bandwidths are suppressed by over a
factor of two. The collapse occurs abruptly at the critical point:
for the middle mode there is virtually no change of dispersion
between $H=0$ and $H=9$~T (Fig.~\ref{Fig4}). Nothing of the sort
happens in the spin-dimer compound TlCuCl$_3$, where the bandwidth
of the two upper excitation branches evolves continuously with
field, and is decreased by only 20\% at $H=12$~T, which is more
than twice $H_c$ \cite{Matsumoto2002}.

The observed phenomenon can hardly be explained by a simple Zeeman
shift of quasiparticle energies. Indeed, the latter is negligible,
as small as $\sim 0.2$~meV between $H_c$ and 11.5~T. Instead, we
suggest that the abrupt spectrum restructuring is is related to
the translational invariance of the ground state wave function for
an AF spin ladder or chain below $H_c$. At $H>H_c$ the emergence
of long-range AF order breaks an {\it additional} discrete
symmetry operation, namely a translation by the structural period
of the ladder. There is no analogue of this in conventional BEC.
Depending on inter-cluster interactions that define the ordering
vector, neither does this necessarily happen in spin cluster
materials. In particular, in TlCuCl$_3$ the induced magnetic
structure retains the periodicity of the underlying crystal
lattice. However, in a uniform AF spin ladder the extra symmetry
breaking is unavoidable, regardless of inter-ladder coupling.

For \IPA\ the spontaneous doubling of the period implies that at
$H>H_c$ the wave vectors $h=0$, $h=0.5$ and $h=1$ all become
equivalent magnetic zone-centers. At the same time, $h=0.25$ and
$h=0.75$ emerge as the new boundaries of the Brillouin zone. The
result is a formation of anticrossing gaps for all magnons at
these wave vectors \cite{Ziman}, where each branch interacts with
its own replica from an adjacent zone. This translates into a
reduction of the zone-boundary energy for the visible (lower)
segments of the two gapped magnons in \IPA. The additional
violated symmetry operation is a microscopic one, and therefore
plays no role in the long-wavelength physics probed at $h=0.5$.

To summarize, any {\it long-wavelength} characteristics of the
field-induced magnetic BEC transition and the magnon condensate,
such as critical indexes, emergence of the Goldstone mode and
behavior of gap energies, appear to be universal. They are not
affected by the 1D topological nature of the normal state in spin
chains and ladders, and are very similar to those in local-cluster
spin systems.  In contrast, the {\it short-wavelength} properties
can be significantly different in these two classes of materials.
In coupled AF spin chains or ladders, unlike in many couple dimer
systems, and unlike in conventional BEC, the transition breaks an
additional discrete symmetry. The result is a radical modification
of the excitation spectrum.

We thank A. Chernyshev (U. of California, Irvine) for his
theoretical insight and to I. Zaliznyak (Brookhaven National
Laboratory) for stressing the significance of the Brilloin zone
folding. Research at ORNL was funded by the United States
Department of Energy, Office of Basic Energy Sciences- Materials
Science, under Contract No. DE-AC05-00OR22725 with UT-Battelle,
LLC. T. M. was partially supported by the US - Japan Cooperative
Research Program on Neutron Scattering between the US DOE and
Japanese MEXT. The work at NIST is supported by the National
Science Foundation under Agreement Nos. DMR-9986442, -0086210, and
-0454672.


\end{document}